## Highlights

**Computational Studies of NaVTe Half Heusler Alloy for Green Energy Applications**

Sumit Kumar,Ashwani Kumar,Anupam,Shyam Lal Gupta,Diwaker

- NaVTe HH alloy is stable in a ferromagnetic phase.
- The electronic band structure shows that it exhibit half-metallic character.
- NaVTe is promising material for optoelectronic applications.
- NaVTe has excellent conductivity at low energy.
- Lattice dynamics studies confirm the dynamic stability of NaVTe

# Computational Studies of NaVTe Half Heusler Alloy for Green Energy Applications

Sumit Kumar[a], Ashwani Kumar[b], Anupam[c], Shyam Lal Gupta[d],* and Diwaker[e],**

[a]*Department of Physics, Government College, Una, 174303, H P , INDIA*
[b]*School of Basic Sciences, Abhilashi University Mandi, Mandi, 175045, H P , INDIA*
[c]*Department of Physics, Rajiv Gandhi Memorial Government College, Jogindernagar, Mandi, 175015, H P , INDIA*
[d]*Department of Physics, HarishChandra Research Institute, Prayagraj, Allahabad, 211019, U P , INDIA*
[e]*Department of Physics, SCVB Government College, Palampur, Kangra, 176061, H P , INDIA*

ABSTRACT

To lessen the quick depletion of fossil fuels and the resulting environmental harm, it is necessary to investigate effective and eco-friendly materials that can convert lost energy into electricity. The structural, optical, electronic, thermoelectric, and thermodynamic properties of the novel half-Heusler material NaVTe were examined in the current work using density functional theory (DFT). The Birch-Murnaghan equations of states were used to confirm the structural stability of the NaVTe HH alloy under investigation. These equations show that the compound in question has structural stability because its ground-state energy levels are negative. For spin-down ($\downarrow$) configurations, NaVTe possesses an energy band gap of 3.2 eV, according to band structure and total density of state analysis. NaVTe is a material that is desirable for optoelectronic applications due to its optical features, which include maximum conductivity and absorption of electromagnetic radiation. The figure of merit and other thermodynamic and thermoelectric parameters are calculated. According to these predicted outcomes, the NaVTe HH alloy would be the ideal option for thermoelectric and renewable energy applications.

## 1. Introduction

Researchers are now interested in alloys made from non-ferromagnetic materials because of the discovery of Heusler type alloys, mainly due to their remarkable thermal and electrical characteristics [1]. There are two varieties of Heusler structures: half and complete Heusler, which are denoted by the letters XYZ and $X_2YZ$, respectively. Here, X and Y stand for transition metal group elements, and Z for main group elements. In some situations, Y might be an alkaline or rare earth element. One characteristic that sets Heusler types apart from other structures is their capacity to change their ferromagnetic characteristics when the constituent elements combine to produce Heusler type crystals [2]. In the field of intermetallic materials, Heusler type compounds or alloys are unique due to their fascinating thermodynamic, electrical, transport, and magnetic properties, which open up a variety of applications [3, 4, 5, 6, 7]. Notably, the quantity of valence electrons is necessary to forecast the basic characteristics of Heusler structures, including superconductivity, electrical, semiconductivity, and magnetic properties [8]. For example, Heusler alloys with 27 valence electrons and no magnetic characteristics exhibit superconducting activity. Full-Heusler compounds with a valence electron count of 24 display semiconductor properties. In addition to being diamagnetic semiconductors, half-Heusler alloys with eight or eighteen valence electrons have exceptional thermoelectric properties [?, 9, 10, 11, 12, 13, 14, 15, 16]. Half-Heusler alloys with the stoichiometry XYZ are introduced, adopting a structure space group F-43m (no: 216). Here, Z is an element from the periodic table's main groups 4 or 5, Y is a transition metal, and X could be an alkali or transition metal. Three face-centered cubic (fcc) lattices with the Wyckoff coordinates (0, 0, 0), (1/2, 1/2, 1/2), and (1/4, 1/4, 1/4) separated by a quarter of the main diagonal make up the Half-Heusler structure [17, 18, 19, 20, 21, 22, 23, 24, 25, 26, 27, 28, 29, 30, 31, 32, 33, 34, 35, 36, 37, 38, 39]. Moradi et al. [40] examined the half-metallic characteristics of the half-Heusler NaZrZ (Z = P, As, Sb) compounds and looked at how pressure affected these characteristics. By replacing the $d^0$ atom of alkaline metals with low-valence transition metals in traditional half-Heuslers, Davatolhagh and associates examined the behavior of half-metallic materials[41]. They demonstrated that the half-metallic property may be produced by the $\beta$ phase of half-Heusler alloys, which are composed of a mixture of $d0$ alkali metals and 3d transition metals. T. Zerrouki et al. [42] used the full potential linear muffin-tin orbital approach in the generalized gradient approximation (GGA) and GGA + U approximation in 2020 to investigate the various physical features of the NbCoSn and NbFeSb half-Heusler compounds. They demonstrated the mechanical stability of the aforementioned compounds and demonstrated the semiconductor nature of the band structure. Using DFT simulations, M. H. Elahmar et al. [43] examined the thermoelectric responses and other characteristics of new half-Heusler NbFe(Mn)Sn compounds and their layered superlattices evolving along [100] direction in 2021. The compounds are primarily ferromagnetic, according to the ground state and magnetic stability characteristics, and the formation energy suggests that the materials under study can be manufactured experimentally. The structural phase stability of the NaVTe HH alloys is examined in the current work. We analyze the theoretical mechanical, dynamic, and thermodynamic stability in the sections that follow. Additionally, the stable structural phase, magnetic and half-metallic characteristics will be evaluated. To our knowledge, there has never been a discussion of the half-metallic characteristics of the NaVTe HH alloy earlier as per literature review.

*Corresponding author
**Corresponding author
 shyamlalgupta@hri.res.in (S.L. Gupta); diwakerphysics@gmail.com ( Diwaker)
ORCID(s): 0000-0003-0472-3436 (S.L. Gupta); 0000-0002-4155-7417 ( Diwaker)

## 2. Computational details

The half-metallic behavior, magnetic, thermodynamical and dynamical structural stability of the NaVTe Heusler alloy were all examined in the current work using density functional theory (DFT). The full potential linearly augmented plane wave (FP-LAPW) basis set in the WIEN2k simulation software program was utilized to successfully solve the Kohn-Sham equations. The Perdew-Burke-Ernzerhof (PBE) exchange-correlation potential computes the generalized gradient approximation, which effectively handles the electron-electron interaction[44, 45, 46, 47]. The generalized gradient approximation (GGA and mBJ) is used to get the detailed electronic structure, and structural stability and magnetic response are successfully examined and analysed. The structural unit cell of NaVTe is divided into two portions to begin the simulations: the interstitial space and the muffin-tin sphere. The Muffin-Tin is one of highly compatible approximation technique that is frequently used to determine the energy state of an electron inside the crystal lattice. The considered values for Muffin-Tin radii (RMT) for atoms Na, V, and Te is 2.23, 2.18, and 2.25, respectively. With a wave vector cut-off of RMT× Kmax=8.0, the electron wave functions beyond the sphere are represented as plane waves, where Kmax is the reciprocal lattice vector. The core as well as the valence electrons are distinguished using a constant energy of magnitude -7.0 Ryd. Whereas, to achieve the self-consistent fieled convergence an energy value of $10^{-5}$ Ryd is successfully used. Using a (6 × 6 × 6) k-point mesh, the Monkhorst pack approach is used to sample the system's Brillouin zone. IRelast [48] in the WIEN2K package is used to find the elastic parameters of NaVTe. Using the Gibbs2 code [49] the thermodynamic properties are investigated. Lattice dynamics studies are performed using phonopy code[50].

## 3. Results and Discussions

This section will cover the different properties of NaVTe half Heusler alloys.

### 3.1. Structural properties

Predicting the physical properties of any material requires first understanding and drawing the crystal structure. The regular Heusler relaxes in centrosymmetric cubic fcc with $L_{21}$ structure and fm-3m space group number. 225, respectively, whereas the half or semi Heusler alloys (HAs) mostly crystallize in non-centrosymmetric cubic fcc with structure type C1b and F-43m with space group no. 216. In fig. 1, these various structures are shown. Type $\alpha$: Na (1/4, 1/4, 1/4), V (0, 0, 0), Te (1/2, 1/2, 1/2); type $\beta$: Na (0, 0, 0), V (1/4, 1/4, 1/4), Te (1/2, 1/2, 1/2); and type $\gamma$: Na (1/2, 1/2, 1/2), V (0, 0, 0), Te (1/4, 1/4, 1/4). These three types of crystal structures are identified by different atomic coordinates and Wyckoff's positions. Table 1 lists the precise Wyckoff's locations for type $\alpha$, type $\beta$ and type $\gamma$ in NaVTe alloy. The computed results exhibited that the phase type $\alpha$ of NaVTe crystal structure is most stable in comparison to $\beta$ and $\gamma$ phase types. Therefore, it is dynamically more stable in the ferromagnetic (FM) phase than the non-magnetic (NM) phase. We observed that the dynamical stability of the proposed system more or less depends upon the atomic positions in the given crystal structure. We have calculated the lowest energy and mean lattice parameters for the cubic NaVTe system with stable type $\alpha$ crystal structure, as shown in Table 2. The Murnaghan equation of the states[27] for volume optimization in the ferromagnetic state is successfully implemented, as expressed in Eqn. 1.

$$E_T(V) = E_o + \frac{B_o V}{B'_o}\left(\frac{(\frac{V_o}{V})^{B'_o}}{B'_o - 1} + 1\right) - \frac{B_o V o}{B'_o - 1} \qquad (1)$$

where $V_0$ represent the unit cell's volume at zero pressure, $B_0$ is the bulk modulus and $E_0$ signifies the minimal total energy. The computed lattice constant ($a^0$) with GGA scheme for NaVTe HH alloy at symmetry alongwith the other parameters $V_0$, $B_0$, and $B'_0$ are provided for the first time in Table 2.

The standard enthalpy of formation gives insight about the thermodynamic stability of the materials under study. It signifies how the material's enthalpy (or heat content) changes when it is created from its constituents. The stability of a Heusler alloy, which typically consists of components arranged in a certain cubic crystal structure, can be determined by calculating its enthalpy of formation in relation to the pure elements. The value of enthalpy of formation $\Delta H_f$ can be known by using

$$\Delta H_f = E_{(NaVTe)} - (E_{Na} + E_V + E_{Te}) \qquad (2)$$

where $E_{(NaVTe)}$ is the enthalpy of formation of HHalloy, $E_{Na}$, $E_V$ and $E_{Te}$ represents the values of enthalpies for constituent atoms. The computed values of $\Delta H_f$ for NaVTe HH alloys is -0.78 eV. Given that negative values suggests that the given HH alloy is stable and most likely can be experimentally synthesized.

### 3.2. Electronic and Magnetic properties

Electronic properties are vital for various memory, smart, and spintronic devices, as they are closely linked to the energy band gap and electronic structure of the alloys. This

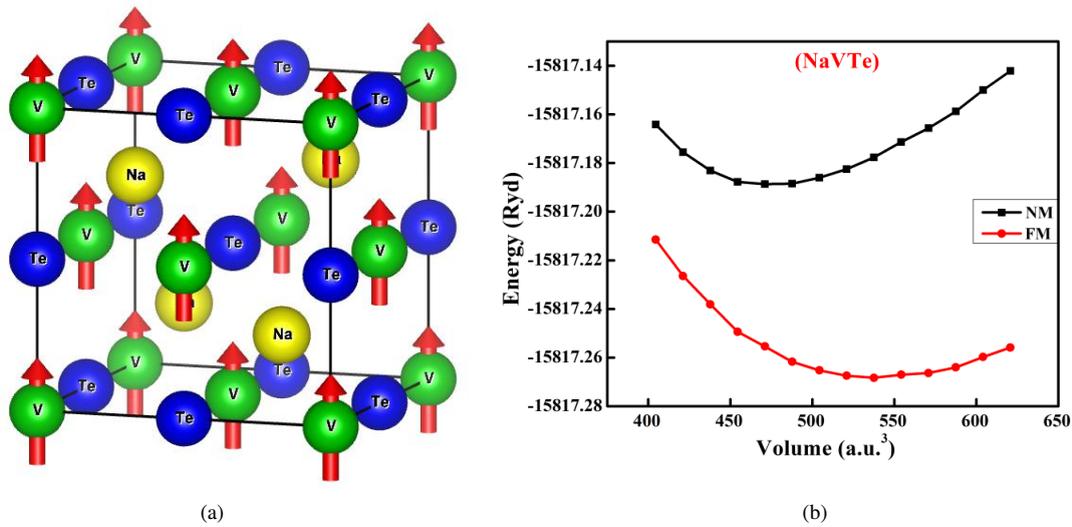

**Figure 1:** [a] Optimized unit cell and [b]Energy versus Volume curve of NaVTe HH alloy

**Table 1**
Wyckoff positions and three potential phase type $\alpha$, $\beta$ and $\gamma$

| Structure | Na | V | Te |
| --- | --- | --- | --- |
| $\alpha$ | (0.25,0.25,0.25) | (0,0,0) | (0.5,0.5,0.5) |
| $\beta$ | (0,0,0) | (0.25,0.25,0.25) | (0.5,0.5,0.5) |
| $\gamma$ | (0.5,0.5,0.5) | (0,0,0) | (0.25,0.25,0.25) |

study investigates the electronic and magnetic properties of the alloy using density of states and band strucutre calculations. We employed the mBJ exchange correlation potential, which effectively predicts band structure and density of states. The electronic band structures for the ferromagnetic (FM) phase within the $C1_b$ structure are shown in fig. 2[a-b]. The density of states (DOS) structures for majority and minority spin channels are presented in fig. 3[a-b]. Fig. 3a displays the total DOS (TDOS) of the alloy, as well as individual atomic DOS. It is evident from the total DOS(TDOS) that there is an asymmetric distribution of majority and minority spin states at fermi level $E_F$, confirming the compound's magnetic nature. Fig. 3b illustrates the partial DOS(PDOS) of constituent three atoms and contribution from d-orbitals from V and Te atoms. Further analysis of TDOS and PDOS reveals that among the Na, V and Te atoms, Vanadium contributed heavily in the formation of spin-states, primarily from the d-states of V and Te. The broadening of states at and above the Fermi level $E_F$ in the majority spin channel can be attributed to the hybridization of d-states of V and Te

An examination of the energy band structure for the majority spin channel (see fig.-2a) reveals a metallic behaviour, characterized by the overlap of valence and conduction bands. In contrast, the minority spin channel(see fig.-2b) exhibits 100% spin polarization at fermi level $E_F$, accompanied by a wide-semiconducting bandgap of approximately 3.2 $eV$ at zone centre $\Gamma$, thereby confirming the half-metallic nature of the alloy under investigation. The energy band structure analysis is in full agreemant with the indication from the PDOS and TDOS analyses. furthermore, the Band structure analysis corroborates the findings from the previous DOS analysis.

As anticipated, the total magnetic moment (Table-3) in the FM $C1_b$ phase ($4.002\mu_B$(mBj)) predominanantly originatess from the V atom (contributing ($3.123\mu_B$), with

**Table 2**
Calculated crystal structure stability parameters.

| Alloy | Type | a0(Å) | B0(GPa) | V (a.u.3) | E(Ry) | | $\Delta H_f$(eV) |
| --- | --- | --- | --- | --- | --- | --- | --- |
| | | FM | FM | FM | FM | NM | |
| NaVTe | $\alpha$ | 6.83 | 35.46 | 539.22 | -15817.27 | -15817.18 | -0.78 |
| | $\beta$ | 6.84 | 34.76 | 538.34 | -15817.21 | -15817.15 | |
| | $\gamma$ | 6.82 | 37.43 | 540.67 | -15817.20 | -15817.12 | |

NaVTe HH alloys for energy applications

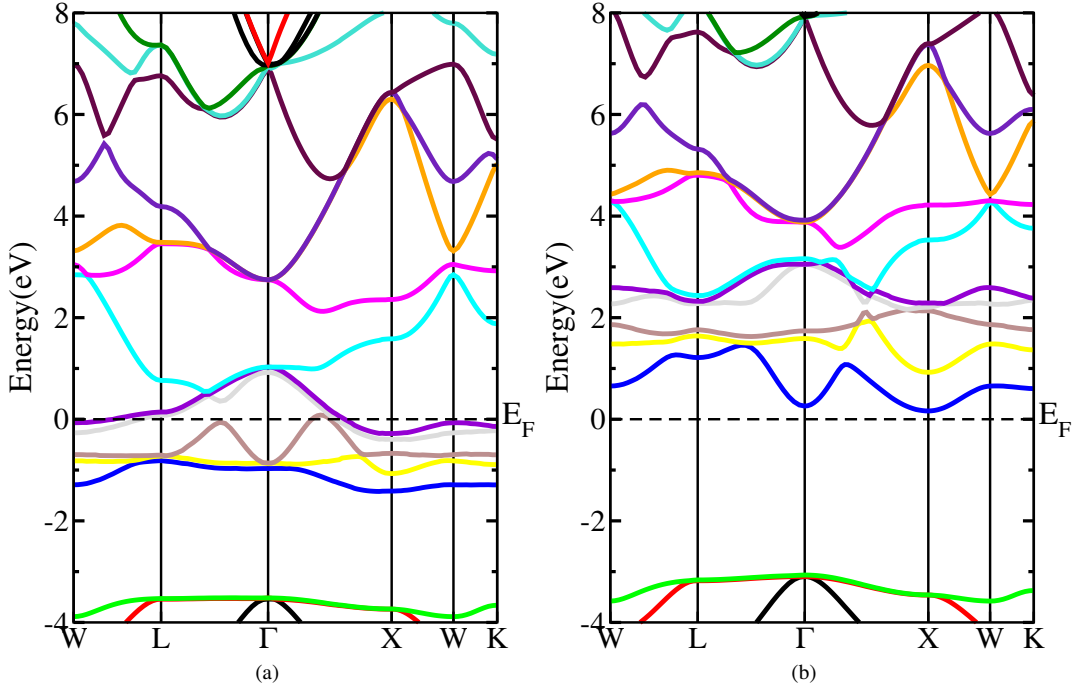

**Figure 2:** Calculated band structure of NaVTe using mBj [a] spin up channel [b] Spin dn channel

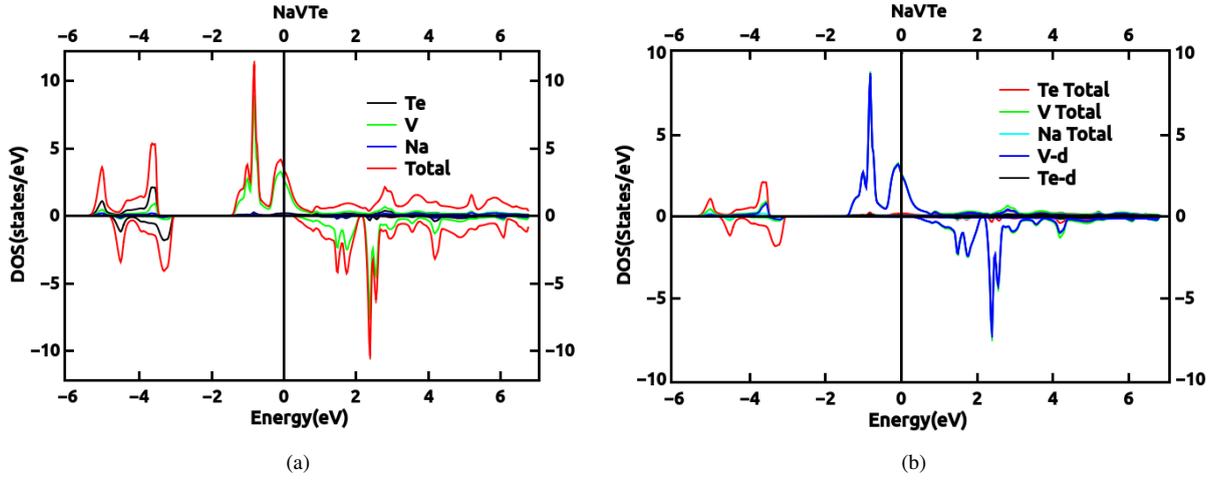

**Figure 3:** [a] Spin-polarized total DOS and [b] atom projected partial DOS of NaVTe employing mBJ

the Te atom exhibiting a negligible induced negative magnetic moment. For the ferromagnetic (FM) phase in the $C1_b$ structure, the total magnetic moment (Table 3) of 4.00155 $\mu_B$ predominantly originates from the V atom, contributing 3.12298 $\mu_B$, whereas the Te atom exhibits a negligible induced negative magnetic moment. This higher contribution of V atom is also evident from the partial Density of States (pDOS) plot (Fig-3a and Fig-3b).

### 3.3. Mechanical stability

By computing the unique elastic parameters and figuring out the relationship between stress and strain, we may ascertain the mechanical and elastic response of NaVTe HH alloy. When exposed to an external force, the alloys' deformation is controlled by the elastic parameters, which enable the material to return to its initial state when the force is withdrawn. It is necessary to comprehend the elastic constants under ideal conditions in order to examine the mechanical stability of alloys. The elastic constants are investigated to ascertain the utilization of alloys in mechanical applications. We can calculate the values of the three distinct elastic parameters, $C_{11}$, $C_{12}$, and $C_{44}$, from the calculated data because both alloys are stable in phase $\alpha$ cubic symmetry. $C_{12}$ measures the transverse deformation, while $C_{11}$ evaluates the longitudinal expansion. The equilibrium elastic coefficients largely meet Born's stability requirements to characterize the mechanical stability which are $C_{11} > 0$, $C_{44} > 0$, $(C_{11}-C_{12})>0$, $(C_{11}+2C_{12})>0$ [29]. Table 3 depicts the mechanical stability and shows that the NaVTe HH alloy meet the stability requirements. Additionally, the two parameters mentioned as bulk modulus (B) and shear modulus (G), respectively determines the stiffness and compressibility [30]. The parameters G and B can be used to calculate the restoring force which comes into play against the fracture, or resistance to plastic deformation[31]. The following equations can be used to get these values The bulk modulus, B is given as

$$B = \frac{c_{11} + 2c_{12}}{3} \quad (3)$$

also, the shear modulus is given as

$$G = \frac{G_R + G_V}{2} \quad (4)$$

Table 3
Value of Magnetic moment for NaVTe per unit cell

| | |
|---|---|
| Total | 4.00155 $\mu_B$ |
| Na | 0.03370 $\mu_B$ |
| V | 3.12298 $\mu_B$ |
| Te | -0.07956 $\mu_B$ |
| Interstitial | 0.92443 $\mu_B$ |

The R and V in the subscript of Eqn. 5 and 6 denotes the Reuss and Voigt bounds and their values are determined by using the subsequent relations given as

$$G_V = \frac{1}{5}(c_{11} - c_{12} + 3c_{44}) \quad (5)$$

and

$$G_R = 5\left[\frac{(c_{11} - c_{12})c_{44}}{3(c_{11} - c_{12}) + 4c_{44}}\right] \quad (6)$$

The mathematical link between the various elastic constants is consistent with the cubic crystal symmetry. The internal strains associated with bond twisting and elongation are computed using the Kleinmann parameter ($\xi$). It assesses how a basic link would bend in the presence of external forces. The NaVTe alloys' increased resistance to a wider range of stresses is indicated by the computed value of ($\xi$), confirming their wider usage in industrial applications. The thermodynamic stability, specific heat at low temperature, and phonon stability were confirmed by calculating the Debye temperature ($\theta_D$) using the average sound velocity. Debye temperature ($\theta_D$) is given as

$$\theta_D = \frac{\hbar}{k_B} 3\sqrt{n * 6\pi^2 \sqrt{V}} \sqrt{\frac{B}{M}} f_v \quad (7)$$

where $k_B$ is the Boltzmann constant, $\hbar$ is Planck's constant, and n is the number of atoms in the primitive cell with volume V unit. A function of the Poisson ratio $\nu$, denoted as $f_\nu$, and M, the compound's mass, correspond to V. For NaVTe HH alloy the predicted Debye temperature ($\theta_D$) at 10GPa pressure is 974.509 K. One important metric for characterizing a material's brittleness or ductility is the Pugh ratio ($\frac{B}{G}$). In general, brittle materials have a ($\frac{B}{G}$) ratio smaller than 1.76. The calculated ($\frac{B}{G}$) ratio values for NaVTe HH alloy show that is brittle in nature. Others parameters which accounts for mechanical stability are also computed and listed in Table 4.

### 3.4. Optical properties

Fig. 4(a) shows the absorption coefficient for NaVTe HH alloy. The absorption coefficient not only describes the amount of light energy absorbed by semiconductors, but it also shows abrupt absorption edges, which are important for figuring out the band gap when impeding energy surpasses it. The incident energy below the band gap prevents electrons from moving from the valence to the conduction band. When plotted against the incident photon frequency range of 0-14 eV the absorption of NaVTe increased noticeably with increasing incident radiation frequency upto 3eV, but remarkably, no absorption occurred when there were no photons on the surfaces of any of these compositions. Based on the plotted graphs, the highest absorption coefficient values for NaVTe HH alloy is close to $1600 \times (10^4/cm)$ and decreases further with an increase in incident frequency range. The calculated results of optical absorption show that these substances are suitable for optoelectronic applications since they can absorb a broad spectrum of ultraviolet and electromagnetic radiations. Reflectivity is the capacity of an outer layer to reflect light. The connection between reflectivity and photon energy is seen in Fig. 4(b). At 3.0 eV, photon energy reaches its maximum value of reflectivity. The static values of R(0) of 0.4 for NaVTe has been observed. The energy loss function L($\omega$) expresses the amount of energy lost by scattering or dispersion during an electron's transition [35]. The scattering probabilities that arise during inner shell transitions serve as the basis for the correlation in Fig. 4(c) which shows the plotting of the optical loss function versus incident photon frequency for NaVTe composition in the range from 0 to 14 eV. From the graphs we find that the peak optical loss values are 1.1 at a photon energy of 9eV for NaVTe HH alloy. The energy loss spectrum indicates that when an external electromagnetic disturbance is applied, the solid reacts. In order to collect optical energy for use in optoelectronic applications, substances that have been identified to have the highest absorption in the UV region where there is less energy loss, reflection, and dispersion can be employed. The electrical conductivity is a representation of the optical current brought on by free carriers produced as a result of incident energy and is denioted by $\sigma[\omega]$. Incoming photon energy excites bound electrons in the valence area, which then move to the conduction band. The graph of optical conductivity versus energy is displaced in Fig. 4(d). The maximum conductivity value recorded for NaVTe HH alloy, is 1.3 (1/fs) at 3 eV photon energy. NaVTe, owing to its excellent photon conductivity at low energy, is a suitable material for optoelectronic applications [36], according to the optical evaluation of NaVTe composition. The numerous peaks detected in the spectra of absorption are signifies the electronic transitions and interactions among the various energy states and also exhibit the complex interactions within electronic structure pertaining to studied material.

**Table 4**
Elastic constants ($C_{ij}$), Bulk, Shear and Young Modulus ($B_V$, $B_R$, B, $G_V$, $G_R$, G and $E_V$, $E_R$, E in GPa), Reuss and Hill Poisson's coefficient($\nu_V$, $\nu_R$ in GPa), Kleinman's parameter ($\zeta$), Transverse, Longitudnal and Average wave velocity ($V_t$, $V_l$ and $V_a$ in m/s), Debye Temperature ($\theta_D$ in K), Pugh's Ratio (k), Chen and Tian Vickers hardness ($H^{C_V}$, $H^{T_V}$ in GPa), Lame's first and second parameter ($\lambda$, $\mu$ in GPa) of NaVTe HH alloy under external stress of 0, 5 and 10 GPa.

| Alloy | NaVTe | | |
|---|---|---|---|
| Stress | 0GPa | 5Gpa | 10GPa |
| Parameters | | | |
| $C_{11}$ | 684.348 | 310.001 | 653.253 |
| $C_{12}$ | -215.996 | -277.238 | -265.368 |
| $C_{11}-C_{12}$ | 900.345 | 587.239 | 918.621 |
| $C_{11}+2C_{12}$ | 252.356 | -244.475 | 122.517 |
| $C_{44}$ | 341.616 | 89.204 | 347.549 |
| $B_V$ | 84.119 | -81.492 | 40.839 |
| $B_R$ | 84.119 | -81.492 | 40.839 |
| B | 84.119 | -81.492 | 40.839 |
| $G_V$ | 385.038 | 170.970 | 392.254 |
| $G_R$ | 378.085 | 123.633 | 385.023 |
| G | 381.562 | 147.302 | 388.638 |
| $E_V$ | 457.331 | 1705.927 | 280.073 |
| $E_R$ | 454.025 | 750.363 | 278.827 |
| E | 455.687 | 1111.771 | 279.454 |
| $\nu_V$ | -0.406 | 3.989 | -0.643 |
| $\nu_R$ | -0.400 | 2.035 | -0.638 |
| $\nu$ | -0.403 | 2.774 | -0.640 |
| $\zeta$ | -0.200 | -0.700 | -0.288 |
| $V_t$ | 9548.955 | 5583.430 | 8655.487 |
| $V_l$ | 11902.887 | 4931.478 | 10380.868 |
| $V_a$ | 10125.310 | 5328.586 | 9102.161 |
| $\theta_D$ | 1009.128 | 553.011 | 974.509 |
| k | 0.220 | -0.553 | 0.105 |
| $H^{C_V}$ | 376.797 | 71.171 | 910.534 |
| $H^{T_V}$ | 345.281 | 242.320 | 812.283 |
| $\lambda$ | -170.256 | -179.693 | -218.253 |
| $\mu$ | 381.562 | 147.302 | 388.638 |

### 3.5. Thermodynamic Properties

For a variety of thermodynamic applications, we have predicted the critical behavior of NaVTe HH alloy using the Quasi-Harmonic Debye model in different pressure and temperature range. In the temperature range of 0-600K and the pressure range of 0-10GPa, the variation of ($C_p$, $C_v$), unit cell volume (V), Debye temperature ($\theta_D$), thermal expansion coefficient ($\alpha$), and entropy change (S) has been studied. The volume change for NaVTe HH alloy with temperature and pressure is displayed in Fig. 5(a) which shows that the volume rises with temperature and rapidly decreases with increase in pressure. The observed behavior thus confirms that pressure has a stronger effect on the volume of alloy under study than temperature. Figures 5(b) and 7(b) illustrate the variation of $C_p$ and $C_v$ vs. T. At low temperatures, we can see that the fluctuations of $C_p$ and $C_v$ are localized for different pressures, indicating a temperature dependence due to anharmonic effects. The specific heat at constant pressure, $C_p$ and $C_v$, reaches a constant value at higher temperatures and adheres to Dulong and Petit's rule. This, as a result of the anharmonic effect which is the same in case of both reported materials at higher temperatures. As a result, we concluded that temperature affect the both $C_p$ and $C_v$ more than pressure. The pressure dependence of entropy for NaVTe HH alloy at different temperatures is displayed in Fig. 6(a). The values of entropies rise with temperature and fall with pressure. It is evident that entropy is more temperature-sensitive than pressure-sensitive. The relationship between $\alpha$ and temperature (T) and pressure (P) is shown schematically in Fig. 7(a). It is noted that $\alpha$ steadily decreases with significant upsurge with P and T values. The value of $\alpha$ is zero at absolute zero temperature, and it rises rapidly with temperature, suggesting that both alloys obey the Debye's $T^3$ law at lower temperatures. The Quasi Harmonic Approximation (QHA) Debye model links $\theta_D$ to a number of solid state physical characteristics, including the melting temperature, heat capacity, and elastic constant. Fig. 6 (b) shows the expected effect of $\theta_D$ on temperature and pressure. The calculated $\theta_D$ at 0K and 0GPa is 245K.

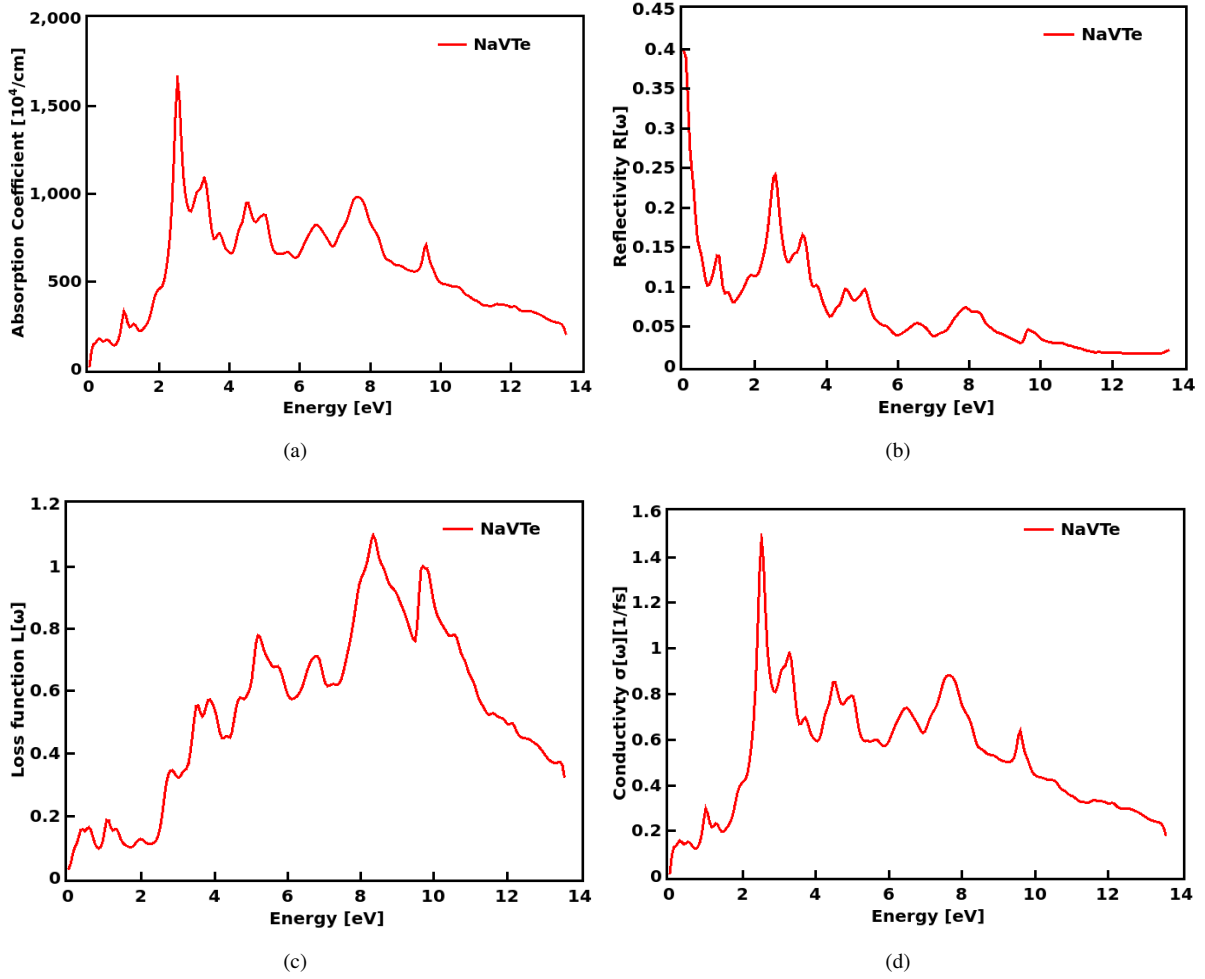

**Figure 4**: Optical properties for NaVTe HH alloy [a]Absorption coefficient, [b]Re flectivity, [c] Loss function, [d] Conductivity

As the temperature rose and the external pressure remained constant, the $\theta_D$ dropped. Furthermore, we observed that the $\theta_D$ increases with increasing pressure at a constant T.

### 3.6. Figure of merit and Lattice dynamic study

The dynamical stability of HH alloy NaVTe has been determined via phonon dispersions in the Brillouin zone as represented in fig. 8. Three atoms make up each of the material's unit cell in terms of crystal structure, which ultimately results in five vibrational modes. Two of these vibrational modes known as acoustic modes are linearly scattered modes at the Γ point that originate from the in-phase atomic displacements of atoms at their basis. The remaining three modes are called optical modes; they originate from the out-of-phase atomic translations and show nearly flat dispersion near the gamma Γ point. In a similar vein, the degeneracy of two sets of optical modes close to the Γ point also clearly displays the crystal symmetry. By closely examining the phonon band structure of materials, we are able to predict their thermodynamic stability with high accuracy [39]. The acoustic phonons, which exhibit great dispersion, are the primary donors to thermal current in the lattice, while the optical modes, which are characterized by low dispersion and group velocity, contribute very little to it. Some materials, such as thermoelectric and thermal insulators, are widely used in a variety of industrial and commercial applications. For best results, they need very low lattice thermal conductivity. We have also computed figure of merit for NaVTe HH alloys as shown in fig. 8b. The figure of merit (ZT) is a basic property for materials that is used to assess their thermoelectric effectiveness. Greater ZT values imply thermoelectric properties that are more effective. The thermal performance (ZT) of the devices is expressed as follows:

$$ZT = S^2 \frac{\sigma T}{K} \qquad (8)$$

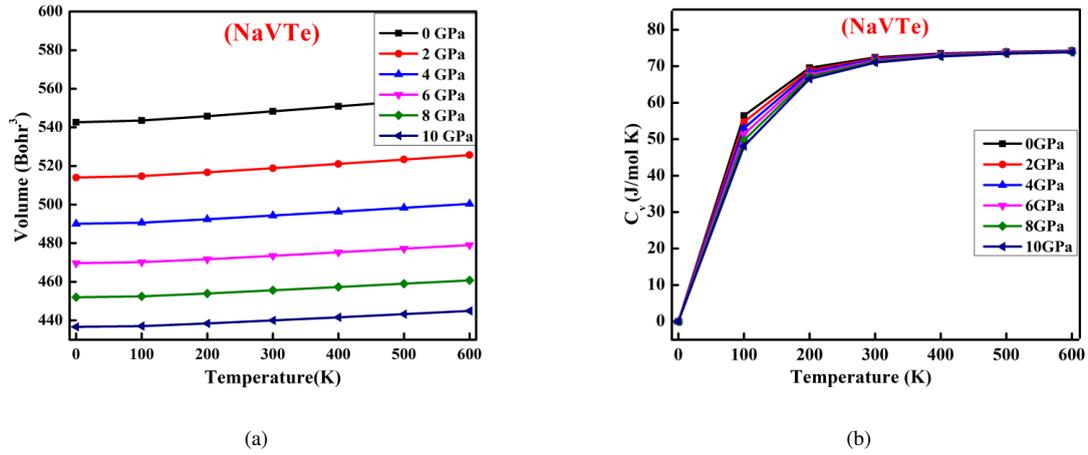

**Figure 5:** [a] Variation of Vol vs T and [b] Variation of $C_v$ vs T for NaVTe HH alloy.

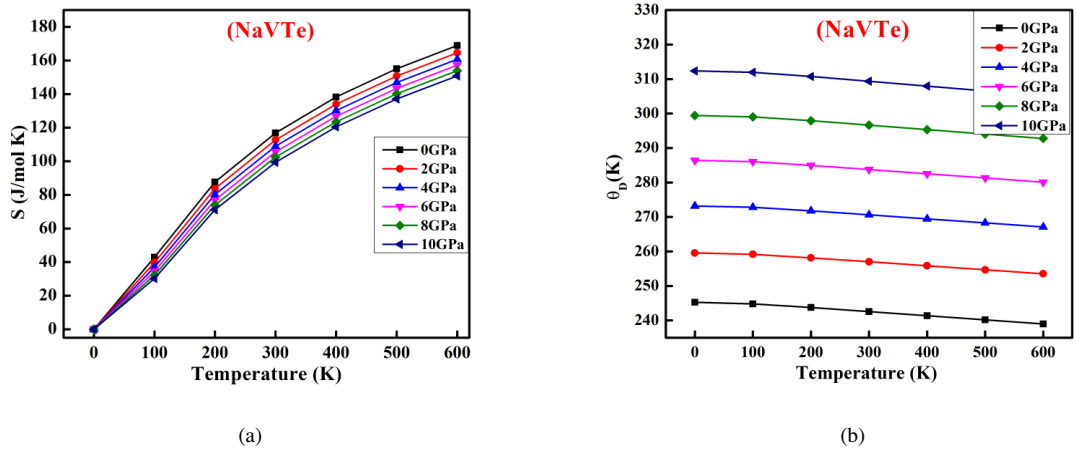

**Figure 6:** [a] Variation of S vs T and [b] Variation of $\theta_D$ vs T for NaVTe HH alloy.

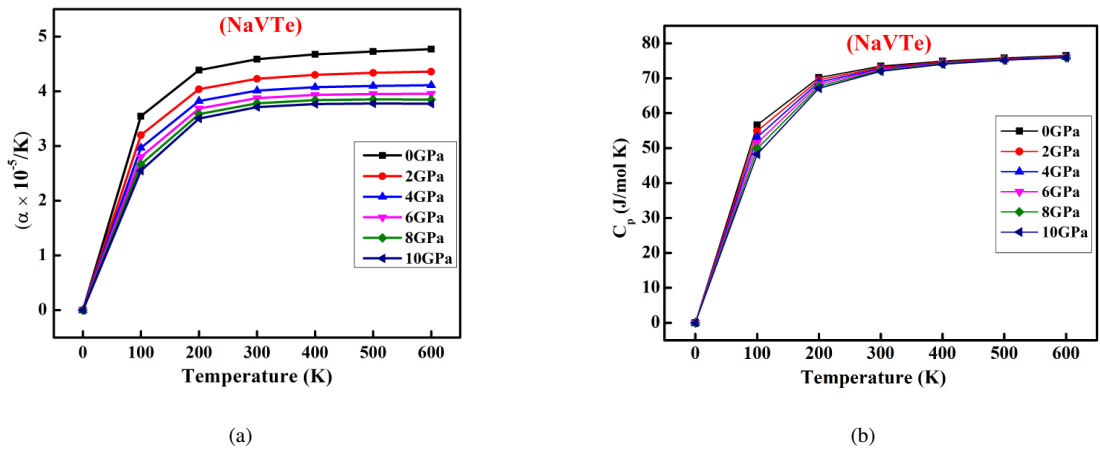

**Figure 7:** [a] Variation of $\alpha$ vs T and [b] Variation of $C_p$ vs T for NaVTe HH alloy..

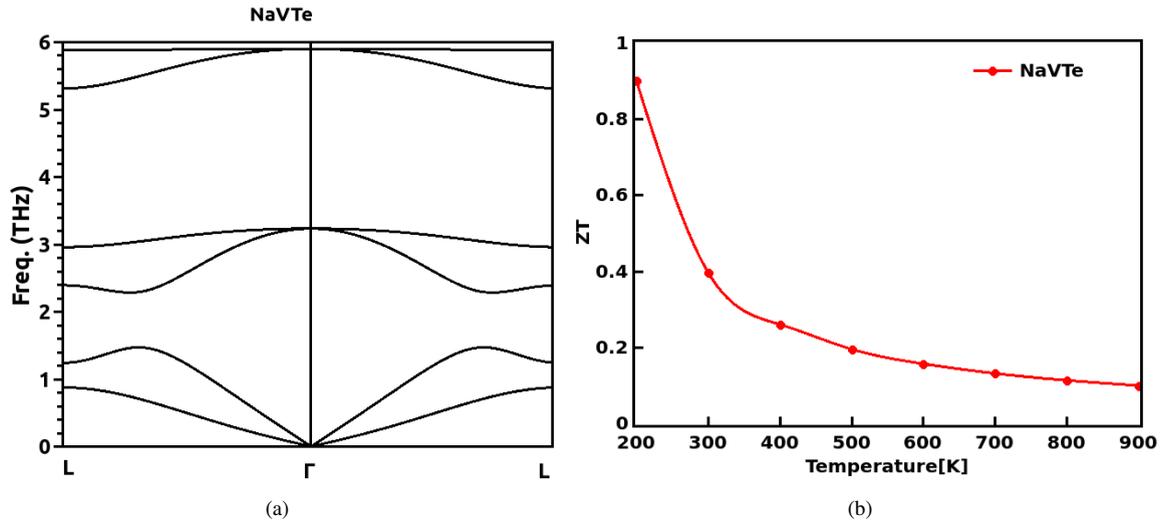

**Figure 8:** Plots of [a] phonon dynamic stability and [b] ZT for NaVTe HH alloys

The values of ZT is 0.9 at 200 K, 0.4 at room temperature and is decreasing with an increase in temperature as depicted in fig. 8b.

## 4. Conclusion

The current study examines novel half-Heusler NaVTe using computational techniques within the framework of DFT. To completely understand the materials' potential for use in green energy applications, a comprehensive analysis was conducted to investigate their structural, optoelectronic, and thermoelectric properties. For the spin down channel, NaVTe shows a half-metallic nature. The results of the partial density of state PDOS show that the contributions of the V atom's d states are responsible for the creation of the conduction and valance bands. In terms of optical characteristics, NaVTe is a promising option for optoelectronic applications due to its optical absorption in the ultraviolet spectrum. Because of its high value figure of merit, NaVTe is a favorable material for thermoelectric applications, according to its thermoelectric properties. It is anticipated that this discovery would encourage scientists and researchers to create thermoelectric and optoelectronic devices based on NaVTe HH Alloy for use in renewable energy applications.

## 5. Declaration of Competing Interest

The authors have no conflict of interest for the work reported in this manuscript.

## CRediT authorship contribution statement

**Sumit Kumar:** Software, Workstation, Data generation. **Ashwani Kumar:** Data compilation, Writing - original draft Writing - review and editing. **Anupam:** Software, Workstation, Data generation. **Shyam Lal Gupta:** Conceptualization, Methodology, Data curation, Writing - original draft Writing - review and editing. **Diwaker:** Conceptualization, Methodology, Data curation, Writing - original draft Writing - review and editing.